\documentclass[10pt,letterpaper]{article}
\usepackage{opex3}
\usepackage{color}
\usepackage{cite}
\usepackage{graphicx}
\usepackage{epstopdf}
\usepackage[breaklinks=true,colorlinks=true,linkcolor=blue,urlcolor=blue,citecolor=blue]{hyperref}
\usepackage{braket}

\begin{document}
\par

\title{Non-classical photon correlation in a two-dimensional photonic lattice}

\author{Jun Gao,$^{1,2}$ Lu-Feng Qiao,$^{1,2}$ Xiao-Feng Lin,$^{1,2}$ Zhi-Qiang Jiao,$^{1,2}$ Zhen Feng,$^{1,2}$ Zheng Zhou,$^{1}$ Zhen-Wei Gao,$^{1}$ Xiao-Yun Xu,$^{1,2}$ Yuan Chen,$^{1,2}$ Hao Tang,$^{1,2}$ and Xian-Min Jin$^{1,2,*}$}

\address{$^1$State Key Laboratory of Advanced Optical Communication Systems and Networks, Department of Physics and Astronomy, Shanghai Jiao Tong University, Shanghai 200240, China\\
$^2$Synergetic Innovation Center of Quantum Information and Quantum Physics, University of Science and Technology of China, Hefei, Anhui 230026, China}
\email{$^*$xianmin.jin@sjtu.edu.cn} 



\begin{abstract}Quantum interference and quantum correlation, as two main features of quantum optics, play an essential role in quantum information applications, such as multi-particle quantum walk and boson sampling. While many experimental demonstrations have been done in one-dimensional waveguide arrays, it remains unexplored in higher dimensions due to tight requirement of manipulating and detecting photons in large-scale. Here, we experimentally observe non-classical correlation of two identical photons in a fully coupled two-dimensional structure, i.e. photonic lattice manufactured by three-dimensional femtosecond laser writing. Photon interference consists of $36$ Hong-Ou-Mandel interference and 9 bunching. The overlap between measured and simulated distribution is up to $0.890\pm0.001$. Clear photon correlation is observed in the two-dimensional photonic lattice. Combining with controllably engineered disorder, our results open new perspectives towards large-scale implementation of quantum simulation on integrated photonic chips.
\end{abstract}



\section{Introduction}
\label{section1}
Quantum systems can be correlated in a way different from classical systems due to unique quantum features, in which quantum interference plays a crucial role \cite{01}. Since the first demonstration of correlations between intensities received by two detectors from a beam of light in 1956, namely the Hanbury Brown and Twiss Effect \cite{02}, an increasing number of quantum correlation phenomena have been studied and experimentally verified \cite{03,04,05,06}. The rapid advance of integrated optics and quantum information makes it possible to manipulate quantum states in a high phase stability as well as highly complex interferometer. Experiments from single-photon generation to multimode interference have been widely demonstrated \cite{07,08,09,10,11,12,13,14}. Photon correlation in one-dimensional structure has been well studied \cite{15,16,17,18,19,20} and finds many applications in quantum information processing and quantum simulation\cite{21,22,23,24,25,26,27,28,29}. Furthermore, prediction of photon correlation under a random unitary matrix, i.e. disordered coupling parameters, has been confirmed exponentially more efficient in experiments than classical computational approaches \cite{30,31,32,33,34,35}. Recently, several progresses attempting to explore structures beyond one dimension have been made, for example, quantum walk in swiss cross structure, elliptic array of waveguides and discrete polarization-time domain \cite{36,37,38}. However, non-classical correlation of identical photons in a genuine spatially two-dimensional structure has never been observed so far.\par
Here, we study the non-classical correlation of two identical photons in a randomly coupled two-dimensional photonic lattices, which is fabricated by femtosecond-laser direct writing. We characterize our photonic chip by using one- and two-photon transmission measurements. We compare the experimental results with the predicted correlation function in both indistinguishable and distinguishable cases and observe non-classical correlation violation up to 20 standard deviations.\\
\section{Theoretical description}
\label{section2}
Typically, photons propagating through evanescently coupled waveguides array can be described by coupled-mode theory. In the Heisenberg picture, the Hamiltonian of the coupled waveguides array is given by
$$H=\sum_{i}^N \beta_i a_i^\dagger a_i + \sum_{i \neq j}^N C_{i,j} a_i^\dagger a_j\eqno{(1)}$$
where $\beta_i$ is propagating constant in waveguide $i$, $C_{i,j}$ is the coupling strength between waveguide $i$ and $j$. When i=j, one can define diagonal entries $C_{i,i}=\beta_i$. For one-dimensional structure, only adjacent waveguides have non-zero coupling strength while two-dimensional system has a much more complex coupling strength matrix with sufficient complexity.

\begin{figure}[htbp]
 \centering
 \includegraphics[width=0.7 \columnwidth]{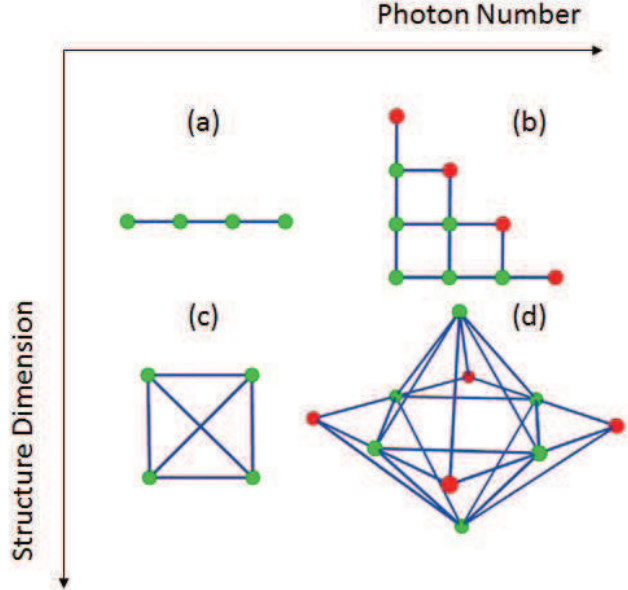}
 \caption{A four-site structure to illustrate the complexity with different photon number and structure dimensions. Each vertices represent the state space of photon populating in the waveguide. (a) and (b) show the graph structure of one-dimensional lattice for one- and two-particle  input respectively. (c) and (d) represent the graph structure of two-dimensional lattice with one- and two-particle injection. Different edges show the possible transition routes between state space. Red nodes highlight all bunching events for two-particle scenarios. }
 \label{Fig.1}
\end{figure}

We will take a four-site structure as an example to illustrate how photon number and structure dimension affect the complexity of the system, as is shown in Fig. \ref{Fig.1}. The basic scenario is one particle evolving on a one-dimensional structure where the size of the Hilbert space coincides with the size of the lattice (see Fig. \ref{Fig.1}(a)). Injecting two particles into a one-dimensional structure can be considered as one particle evolving on a two-dimensional graph. Fig. \ref{Fig.1}(b) shows a larger state space even though physical system is still a four-site linear chain as shown in Fig.\ref{Fig.1}(a). Things become different when we take a spatially two-dimensional structure into consideration (see Fig. \ref{Fig.1}(c)). It cannot be treated equivalent to the case where two particles evolving on a one-dimensional graph. The number of state space equals to site number of the lattice, however, due to full coupling between different sites, a two-dimensional structure do show more complexity than one-dimensional ones, namely, with off-near-diagonal terms in Hamiltonian matrix. Furthermore, the power of particle number and structure dimension can be applied simultaneously. As we can see, Fig. \ref{Fig.1}(d) has a much larger complexity and connectivity.

The evolution of photonic creation (annihilation) operator is determined by Heisenberg equation
$$i \frac{da_{i}^\dagger(t)}{dt}+\sum_{j}^N C_{i,j} a_j^\dagger(t)=0\eqno{(2)}$$
By integrating with substitution $z=ct$, one obtains
$$a_i^\dagger(z)=\sum_{i}^N U_{ij}(z)a_j^\dagger(z=0)\eqno{(3)}$$
where $z$ is the propagation length and $U_{ij}(z)=({e^{iCz}})_{ij}$ is the unitary evolution operator along $z$ direction. The initial input states for one- and two-photon can be written as
$$\ket{\varphi_0}=a_i^\dagger\ket{0}\eqno{(4)}$$
$$\ket{\psi_0}=a_i^\dagger a_j^\dagger\ket{0}\ket{0}\eqno{(5)}$$
respectively, where $\ket{0}$ is the vacuum state and $i,j$ indicate the waveguides where photons are injected.

To observe single-photon distribution and two-photon correlation, we shall consider average photon number in the waveguide. For single-photon case,
$$\braket{n}_{i^\prime}=\braket{a_{i^\prime}^\dagger(z) a_{i^\prime}(z)}_{\varphi_0}=|U_{i^\prime i}|^2\eqno{(6)}$$
Similarly, two-photon correlation function can be defined as
$$\Gamma_{i^\prime j^\prime}=\braket{a_{i^{\prime}}^\dagger (z)a_{j^{\prime}}^\dagger (z)a_{i^{\prime}}(z)a_{j^{\prime}}(z)}_{\psi_0}=\frac{1}{1+\delta_{i^\prime j^\prime}}|U_{i^\prime i}U_{j^\prime j}+U_{i^\prime j}U_{j^\prime i}|^2\eqno{(7)}$$
Delta function is introduced to eliminate the coefficient $\sqrt{2}$ of the bunching events. As for classical correlation function,
$$\Gamma_{i^\prime j^\prime}^\prime=\frac{1}{1+\delta_{i^\prime j^\prime}}(|U_{i^\prime i}U_{j^\prime j}|^2+|U_{i^\prime j}U_{j^\prime i}|^2)\eqno{(8)}$$
This represents the probability to detect one photon at output ${i^\prime}$ while another photon at output ${j^\prime}$.
Violation between quantum correlation and classical correlation, according to reference \cite{15,21}, can be defined as
$$V_{ij}=\frac{2}{3}\sqrt{\Gamma_{ii}\Gamma_{jj}}-\Gamma_{ij}<0\eqno{(9)}$$
$V_{ij}$ is found to be positive only when quantum interference happens,which quantifies the nature of non-classical correlation.

However, in our case, the coupling length of each waveguide in the coupling zone is randomly chosen to introduce disorder. Therefore, the coupling strength and evolution time of each waveguide is not uniform and it becomes nearly impossible to determine every term of Hamiltonian, which leads to difficulties for theoretical analysis of such dynamics process. Hence it is more desirable to make direct characterization of the evolution operator utilizing single- and two-photon interference procedure \cite{39}.\\

\section{Experimental demonstration}
\label{section3}
\begin{figure}
 \centering
 \includegraphics[width=1\columnwidth]{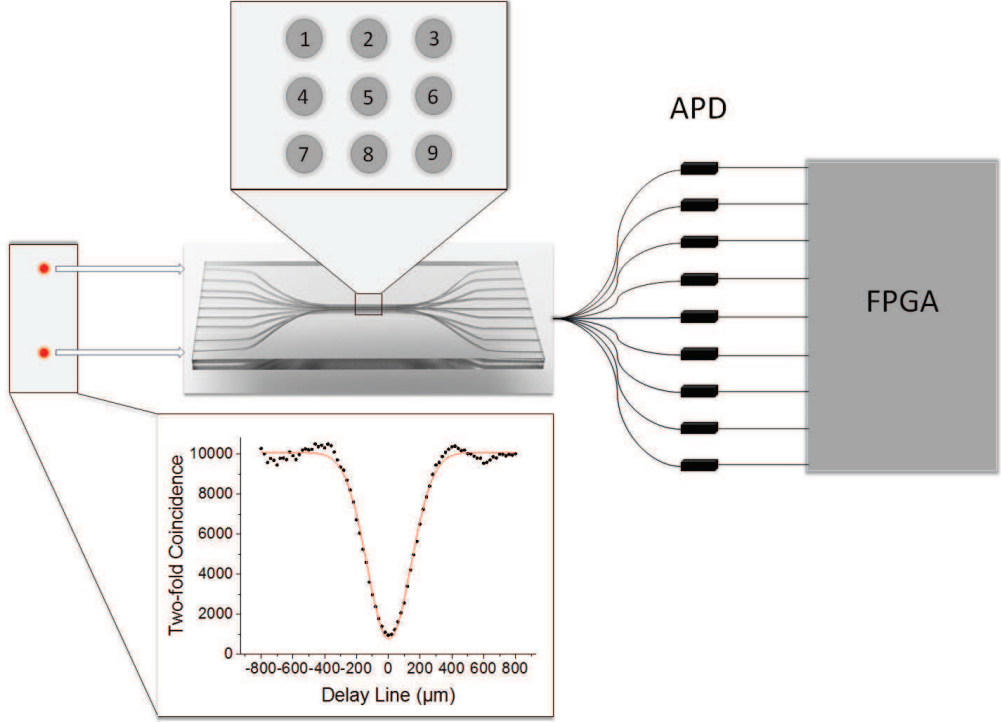}
 \caption{The experiment setup can be divided into three parts. One-photon and two-photon states generated by degenerate spontaneous parameter downconversion are injected into the photonic chip with temporal delay lines to synchronize the two photons. The Hong-Ou-Mandel dip shows the distinguishability of the two photons. The circuit consists of $9$ modes with a spatially two-dimensional square structure. Coincidence detection is performed at the output of the chip using 9 avalanche photodetectors (APD) and a homemade field-programmable gate array (FPGA).}
\label{Fig.2}
\end{figure}

Our experimental setup is shown schematically in Fig. \ref{Fig.2} Two identical photons are generated via spontaneous parameter downconversion (SPDC). The ultraviolet pump pulses are frequency doubled from a mode-locked Ti:Sapphire oscillator. The central wavelength of the mode-locked oscillator is $780nm$ and the pulse width is $130fs$ with a repetition rate of $76MHz$. The UV pulses are then focused on a $2mm$ beta-barium-borate (BBO) crystal cut for type-II non-collinear degenerate SPDC \cite{40}. The power of the UV pulses is $600mW$ in order to produce highly bright SPDC photons as well as acceptable multi-photon emission. The identical photon pair is obtained by steering SPDC photons through polarized beam splitters (PBS) where polarization entanglement is released.

To achieve high spectral indistinguishability, the SPDC photons are filtered by $3$-$nm$ bandpass filters. To characterize the distinguishability of two photons, we perform Hong-Ou-Mandel (HOM) interference by feeding two photons into a fiber beam splitter while scanning a temporal delay via a motorized linear stage. The purity of two photons can be bounded by HOM interference visibility. The raw visibility of HOM dip is $92.4\pm0.4\%$ under the condition of imperfect beam splitter. The measured splitting ratio $47:53$ implies a photon indistinguishability up to $98.1\%$ (see Figure. \ref{Fig.2} inset).

\begin{figure}[htbp]
 \centering
 \includegraphics[width=1 \columnwidth]{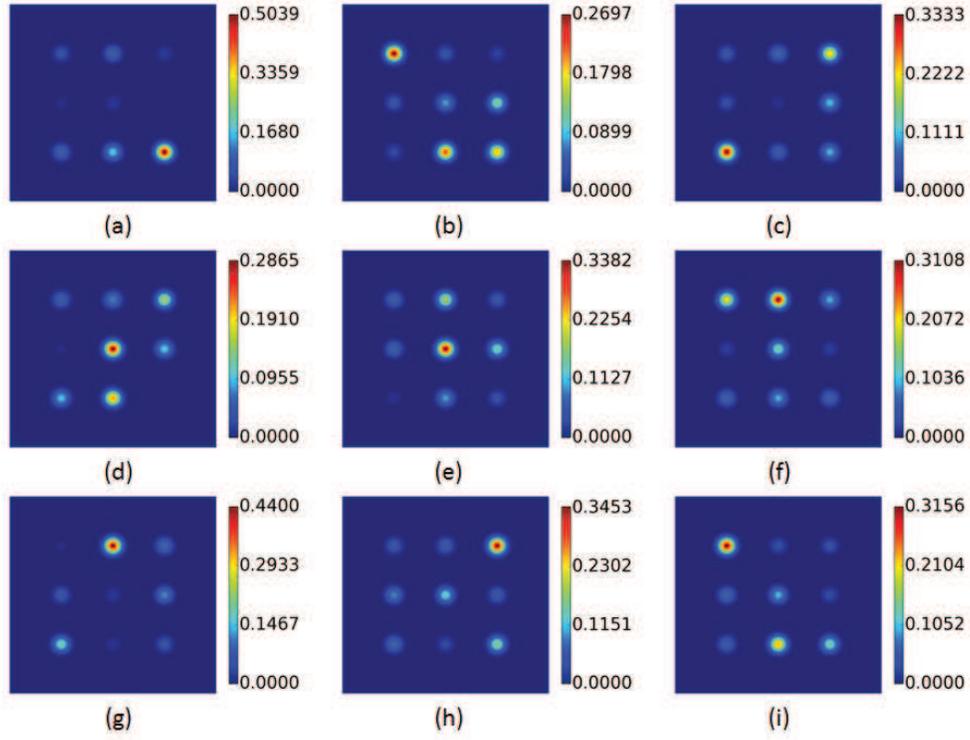}
 \caption{(a)-(i) show the measurement of single-photon distribution at the output end of the chip when injecting port ranges from 1 to 9. These irregular results verify the random coupling among different waveguides. Results also show few zero terms, indicating the structure is fully coupled. Compared to one-dimensional scenario, two-dimensional structure allows photon coupling to further site without long time propagation.}
 \label{Fig.3}
\end{figure}

The two-dimensional photonic lattice is fabricated on a pure borosilicate chip via fs-laser direct writing \cite{41}. A high power laser pulse at the wavelength of $513nm$ is focused at $170um$ underneath the surface of a pure borosilicate glass to obtain a permanent and localized refractive index change. During the fabrication process, the translation stage holding the glass chip can move in three directions, and transverse tolerance is kept less than $100nm$, which enable us to fabricate waveguide arrays in three dimensions. The two-dimensional photonic lattice is formed on the cross section perpendicular to the propagating direction. Each waveguide represent a lattice by trapping photons in an individual site. The propagating length allows us to set evolution and coupling strength between sites. 

In our case, we fabricate a three-layer square two-dimensional structure consisting of nine waveguides. The three-layer structure is adiabatically transformed into one-layer at both end of the chip to match one-dimensionally aligned fiber array in V-groove.  Distance between each waveguide is close enough to guarantee strong evanescent coupling. The disorder is introduced by designing and engineering a differential coupling length. Both photons are prepared in vertical polarization and coupled into the polarization maintaining (PM) fibers in V-groove. The photons then are guided to Avalanche photodiodes (APDs) through another V-groove fiber array butt-coupled to the chip. All coincidence events are recorded by a homemade FPGA system.

In our experiment, the writing speed of our fabrication is 10 mm/s. We are therefore able to complete the writing of 9 waveguides within two minutes. The change of parameters of laser and environment is negligible in such short time, which guarantees a uniform loss for all waveguides under same parameters. In three-dimensional coupling zone, every waveguide is written to be straight with a uniform loss of $0.2 dB/cm$. In the zone where we adiabatically transform 9 waveguides from three dimensions to two dimensions, we introduce differential bending associated with loss. However, this differential loss can be attributed into the relative coupling parameters between waveguides \cite{31}. 

As mentioned at the end of Section. \ref{section2}, we directly characterize the photonic chip following reference 39. We first use single photon to measure the moduli of this complex-valued matrix. The result is shown in Fig. \ref{Fig.3} and suggests that different waveguides are randomly coupled. The result can also be regarded as single-photon output distribution from different input ports. We can see that all waveguides have been fully coupled and the output distribution is highly asymmetric which again verifies the randomness in the coupling zone.

In order to determine relative phases, we performed a series of HOM-dip scanning procedure between different modes. The relative phases can be retrieved by their corresponding HOM visibilities. Notice that it is not necessary to characterize the entire system, alternatively, a submatrix of parameters in which only input modes get involved will be sufficient. In total, we have done $24$ HOM-dip scans with input of $1\&9$, $1\&8$ and $8\&9$, which take $11$ hours for each pair.

\begin{figure}[htbp]
 \centering
 \includegraphics[width=1\columnwidth]{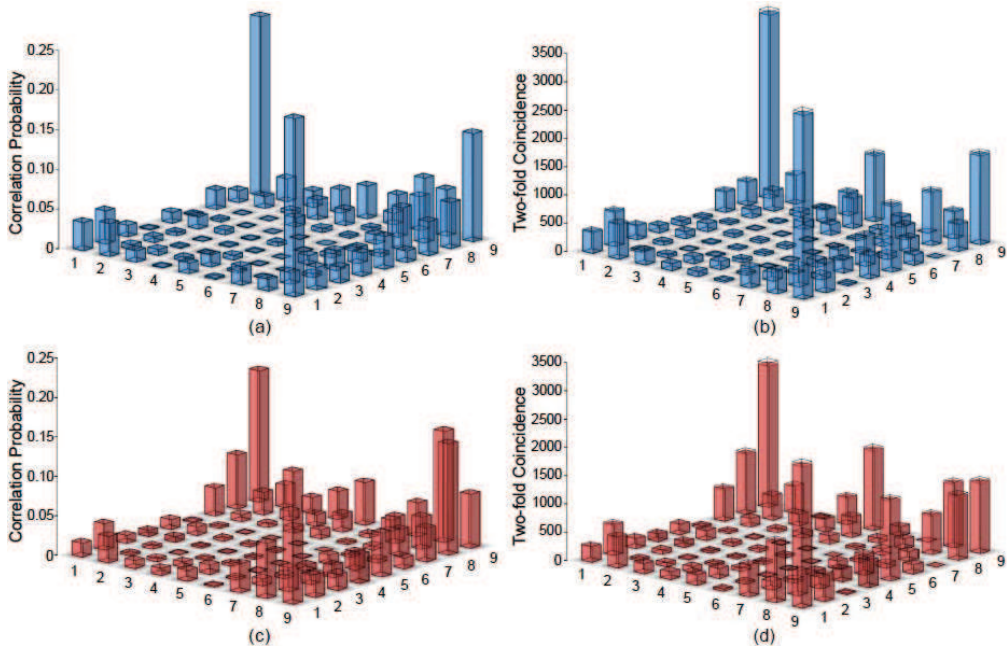}
 \caption{Photon correlation function when two Indistinguishable (blue) and distinguishable (red) photons are coupled to waveguide $1\&9$. Distinguishability is introduced by a controllable temporal delay. (a) and (c) are simulated results while (b) and (d) are measured data. Photon bunching events are measured by photon number resolving detection with a 50:50 fiber beam splitter. Each data set is collected for 10 minutes. Transparent histograms indicate the upper bound of one standard deviation.}
 \label{Fig.4}
\end{figure}

Photon correlation is measured by injecting two photons into mode $1$ and mode $9$ and detecting $2$-fold coincident events of all combination of outputs. Two-photon bunching in one waveguide is detected by using photon number resolving detection with a 50:50 fiber beam splitter. By tuning the temporal delay between two injected photons, we are able to observe both classical and non-classical correlation. The simulation results are shown in Fig. \ref{Fig.4}(a) and \ref{Fig.4}(c), and the experimental results are shown in Fig. \ref{Fig.4}(b) and \ref{Fig.4}(d). Identical photon interference shows more photon bunching events due to bosonic nature compared to distinguishable scenario. Localization phenomenon is also observed, which we attribute to the localization of single photons induced by the static disorder we introduced, see Fig. \ref{Fig.3}(a) and \ref{Fig.3}(i). As for non-classical correlation analysis, we will further discuss this in Section. \ref{section4}.\\

\section{Discussion}
\label{section4}
We define a similarity function to quantify the overlap between experimental measured and ideal simulated distribution as following
$$S=\frac{(\sum_{i,j}^N \sqrt{\Gamma_{ij}\Gamma_{ij}^\prime})^2}{\sum_{i,j}^N \Gamma_{ij}\sum_{i,j}^N \Gamma_{ij}^\prime}\eqno{(10)}$$
$S=0.890\pm0.001$ and $S=0.916\pm0.001$ for identical and temporal delayed photons. Overlap mismatch can be attributed to three factors. The first one is spectral distinguishability. This can be overcome by injecting frequency-uncorrelated photons \cite{42} or by using a narrower bandpass filter at the cost of source brightness. The second factor is loss, which is dominant in the state of arts of integrated photonic circuit. Theoretical analysis shows that losses before circuit do not affect the post-selected results. However, unbalanced losses at the output, namely differential coupling efficiency to fibers, do affect the post-selected results \cite{31}. This can be solved by improving the facet coupling efficiency with dedicated alignment or a multi-mode V-groove fiber array. The last factor is the intrinsic high-order emission from SPDC source. Heralded single photon or single emitter source will help to improve the performance \cite{43,44}.

\begin{figure}[htbp]
 \centering
 \includegraphics[width=0.9 \columnwidth]{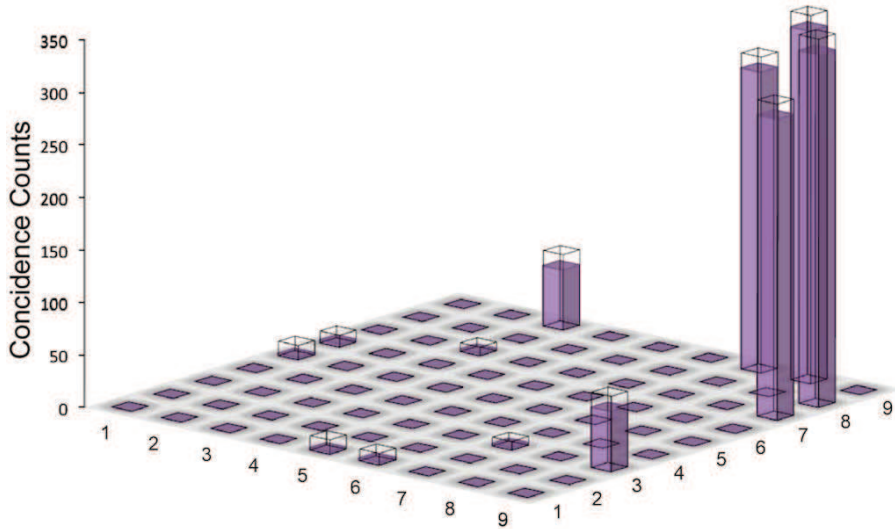}
 \caption{Amount of violations of Cauchy-Schwarz inequality from experimental data. All negative terms have been ignored. Transparent histograms indicate upper bound of one standard deviation. A maximum violation up to 20 standard deviation has been observed.}
 \label{Fig.5}
\end{figure}

To quantify the non-classical correlation in spatial domain, we calculate the violation of Cauchy-Schwarz inequality discussed in Section. \ref{section2}. The violation of this inequality is a sign of photon bunching events which only occur in quantum regime. The result is plotted in Fig. \ref{Fig.5}. All negative terms (no violation) have been ignored. Assuming photon number statistics obey Poissonian statistics, we obtain a maximum violation of classical limit with $20$ standard deviations, which is a strong evidence of non-classical quantum behavior.\\

\section{Conclusion}
\label{section5}
In conclusion, we have experimentally studied the quantum correlation of two identical photons in a fully coupled genuine two-dimensional photonic lattice. We also introduce disorder by randomizing the coupling length for each waveguide, which adds a new capacity into our two-dimensional evolution system for future exploration beyond standard quantum random walk. A good match between measured and simulated results again verifies the capacity of robust and precise on-chip manipulation of quantum states at single-photon level. Strong correlation reveals that geometric two-dimensional structure is of fundamental interest and has many intrigue properties to explore. Our three-dimensionally fabricated photonic chip associated with multi-photon-chip interface may serve as a good candidate in exploring quantum simulation in a more complex regime.\\

\section*{Acknowledgment}
The authors thank J.-W. Pan for helpful discussions. This research leading to the results reported here was supported by the National Natural Science Foundation of China under Grant No.11374211, the Innovation Program of Shanghai Municipal Education Commission (No.14ZZ020), Shanghai Science and Technology Development Funds (No.15QA1402200), and the open fund from HPCL (No.201511-01). X.-M.J. acknowledges support from the National Young 1000 Talents Plan.\\

\end{document}